\documentclass[english]{cfasta}
\usepackage{booktabs}
\usepackage{multirow}
\usepackage{threeparttable}
\usepackage{makecell}
\usepackage[comma,numbers,square,sort&compress]{natbib}
\usepackage{hyperref}
\usepackage{epstopdf}
\usepackage{verbatim}
\usepackage{gbt7714new}
\usepackage{amsfonts,amssymb,amsmath}
\usepackage{bm}
\usepackage[skip=5pt]{caption}
\begin{document}

\title{Distributed Flocking Control of Aerial Vehicles Based on a Markov Random Field}



\author{Guobin Zhu,
        Shanwei Fan,
        Qingrui Zhang$^{*}$ }
\affiliation{School of Aeronautics and Astronautics, Sun Yat-sen University, Shenzhen 518107, P.R. China
    \email{zhugb@mail2.sysu.edu.cn, fanshw3@mail2.sysu.edu.cn, zhangqr9@mail.sysu.edu.cn}}

\maketitle

\begin{abstract}
The distributed flocking control of collective aerial vehicles has extraordinary advantages in scalability and reliability, \emph{etc.} However, it is still challenging to design a reliable, efficient, and responsive flocking algorithm. In this paper, a distributed predictive flocking framework is presented based on a Markov random field (MRF). The MRF is used to characterize the optimization problem that is eventually resolved by discretizing the input space. Potential functions are employed to describe the interactions between aerial vehicles and as indicators of flight performance. The dynamic constraints are taken into account in the candidate feasible trajectories which correspond to random variables. Numerical simulation shows that compared with some existing latest methods, the proposed algorithm has better-flocking cohesion and control efficiency performances. Experiments are also conducted to demonstrate the feasibility of the proposed algorithm.
\end{abstract}


\footnotetext{This work is supported in part by the National Nature Science Foundation of China under Grant 62103451 and Shenzhen Science and Technology Program JCYJ20220530145209021 ($^{*}$Corresponding author).}

\section{Introduction}\label{sec:Introduction}
In recent years, flocking control of multi-UAV (MUV) systems has attracted significant attention from scholars due to its diverse application backgrounds, such as rescue, detection, transportation, and mapping, \emph{etc.} \cite{Hu2021TRO,zhang2017JA, Zhang2021JGCD}. A great amount of research \cite{reynolds_flocks_1987-1,Vicsek2017PR,fang_flocking_2017,ma_o-flocking_2020} has been devoted to the fundamental goals of flocking control, including controlling large gatherings of UAVs and consistent movement, with little regard for flight performance. Furthermore, the MUV system is characterized by an enormous amount of individuals with limited physical resources, such as a lack of computational resources. It is rarely addressed how to ensure the efficient execution of swarm tasks. Therefore, it is still a significant challenge to develop a feasible, efficient, scalable flocking control algorithm with outstanding performance.

In \cite{reynolds_flocks_1987-1}, Reynolds first proposed the flocking model from the perspective of animation production, which includes three principles: \emph{repulsion}, \emph{attraction}, and \emph{alignment}, to recreate the movement of the birds ideally. \emph{Repulsion} ensures collision avoidance between individuals, \emph{attraction} promotes proximity through tilting each individual toward the local center of mass and \emph{alignment} achieves the consistency of group velocity. Then, a variety of flocking control methods following Reynolds criteria have been proposed \cite{olfati-saber_flocking_2006,Vasarhelyi2018SR,fernando_online_2021}. Olfati-saber \cite{olfati-saber_flocking_2006} stated a general design framework to address the aggregation and consistency problems. V\'as\'arhelyi \cite{Vasarhelyi2018SR} offered a new braking curve and improved velocity alignment performance. Fernando \cite{fernando_online_2021} screened the future states using discrete control space and could also realize collective behavior. By tackling an optimization with multi-constraint to determine the optimal control input, Lyu \cite{lyu_multivehicle_2021} achieved the fundamental goals of flocking control, \emph{etc.} However, due to the complexity and changeability of the task objectives and real scenarios, These methods, embedded with Reynolds criteria, do not consider how to improve flight performance and are insufficiently practical.
\begin{figure}[!htb]
    \centering
    \includegraphics[width=1\linewidth]{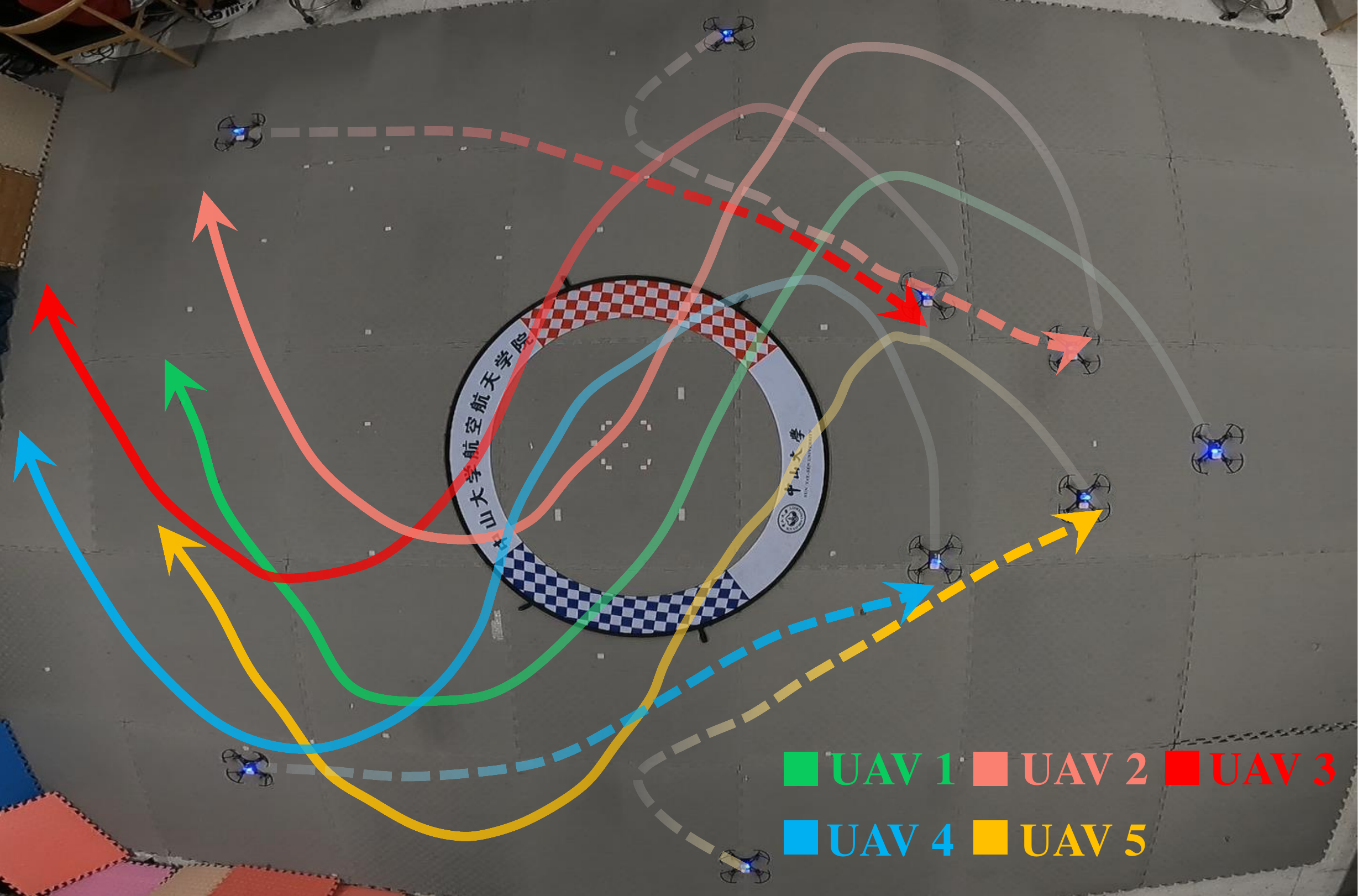}
    \caption{4 UAVs moved from the initial position and approached the leader UAV 1 (Dotted line). After forming a flock, the 5 UAVs move along an S-shaped trajectory, while keeping the distance between each other basically unchanged (Solid line).}
    \label{fig:real_trajectory}
\end{figure}

Artificial potential fields (APFs) have been shown to be effective in the above methods due to their simplicity and versatility. In general, the APFs simulate the attraction and repulsion force in the gravitational field. Attraction can direct each UAV to move toward the target or maintain formation, whereas repulsion can prevent collisions. However, such an attractive and repulsive effect is determined by the gradient descending direction of the potential function and reflects a passive characteristic, which may cause an oscillation \cite{soria2021Nature}. Unnatural oscillations are more noticeable with a variety of motion patterns.

In summary, to solve the problem of insufficient responsiveness and oscillatory flight performance, a predictive flocking framework is proposed in this paper, which is motivated by the predictive intelligence of natural bio-individuals \cite{montague_bee_1995}. The dynamic constraints are used to predict the future states thus there is no problem of dynamics being unfeasible, and the potential functions are regarded as the evaluation criterion to screen predicted states. Then, the optimal control input of the UAV corresponds to the most acceptable candidate trajectory by the fact that the system's expected state changes in the direction of the energy reduction. To represent the possibility of each state to be examined, the Markov random field (MRF) \cite{fernando_online_2021,rezeck_flocking-segregative_2021} is introduced, where the interrelations among UAVs can be treated by the joint probability distribution of the random variables. In this way, selecting the optimal control input is transformed into solving the maximum joint probability density. Such an approach can conveniently address flight performance through several potential functions. And its control effect is also validated via the experimental analysis.

The rest of the paper is organized as follows. Section \ref{sec:Problem Formulation} formulates the problem of interest. Section \ref{sec:Approach} presents the main algorithm. Simulation and experiment results are provided in Section \ref{sec:Experiment}. Conclusions are given in Section \ref{sec:Conclusion}.

\section{Problem Formulation}\label{sec:Problem Formulation}
\begin{figure}[!htb]
    \centering
    \includegraphics[width=0.93\linewidth]{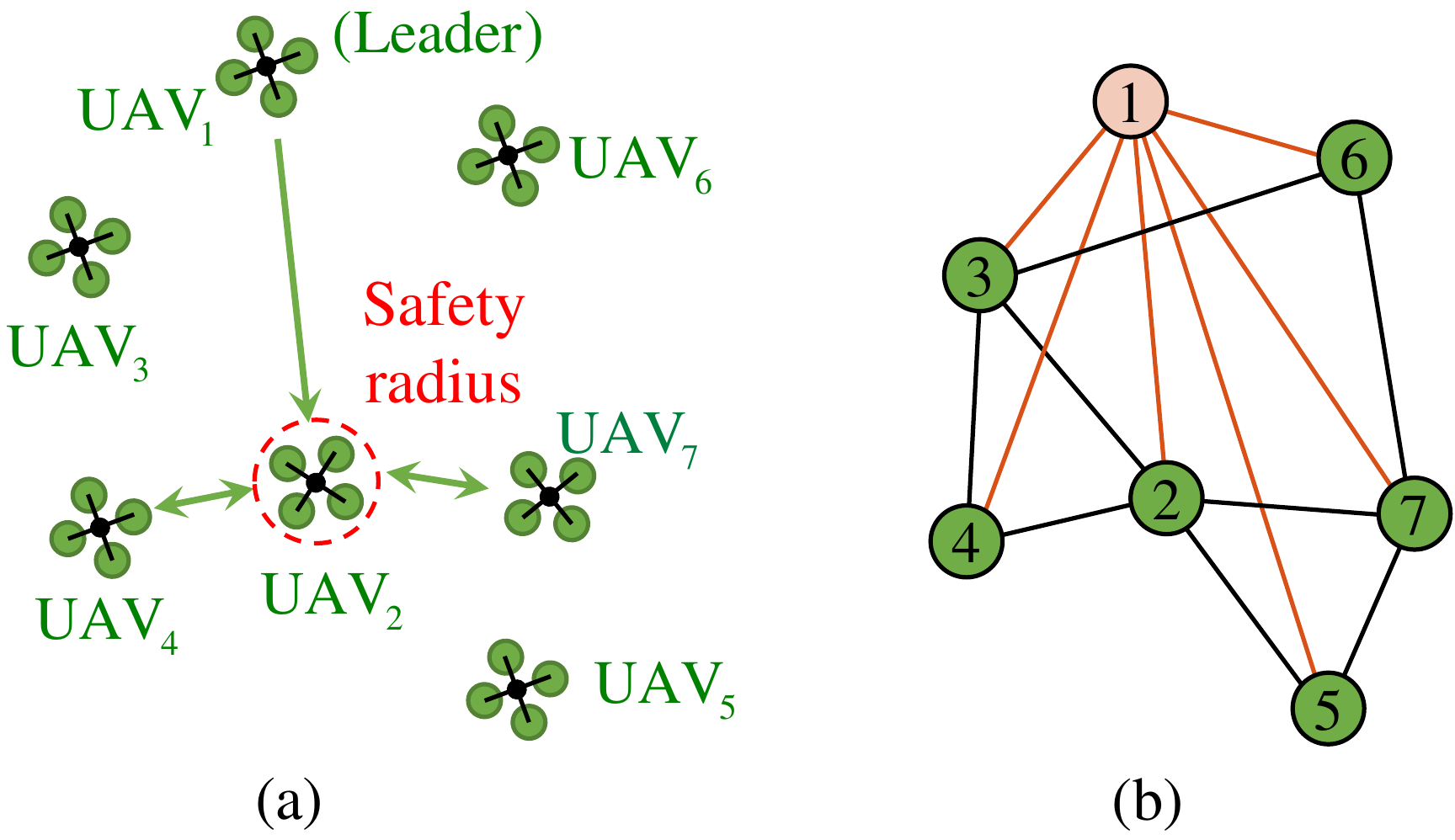}
    \caption{Interactions among UAVs. (a) The red circle represents the UAV's safety radius $r_{coll}$, so $\Vert\mathbf{p}_{ij}\Vert \geq 2r_{coll}, \forall j \in \mathcal{N}_i$ is required. The green line with two arrows implies that the two UAVs on each side could communicate with each other, while the one with one arrow indicates that a UAV has access to the leader. (b) The set of all nodes that are directly connected is a clique. The orange line indicates that the leader's information is transmitted to the follower.}
    \label{fig:UAV_connection}
\end{figure}

There is a set of UAVs with $N$ members, represented by $\mathcal{A}=\left \{1,2,\ldots,N \right \}$.  For each UAV $i \in \mathcal{A}$, its state consists of the position $\mathbf{p}_i \in \mathbb{R} ^{m}$ and its $(n - 1)$-th derivative, where $m=2,3$, \emph{i.e.}, $\mathbf{x}_i= [{\mathbf{p}}^T_i, {\dot{\mathbf{p}}}^T_i, \ldots,$ $ {\mathbf{p} ^{(n-1)}}^T_i]^T \in \mathbb{R} ^{nm}$. Let $\mathbf{u}=\mathbf{p}^{(n)}$ be the control input of. Considering differential flatness, the dynamic constraint can be expressed as \cite{fernando_online_2021}.
\begin{equation}\label{eq:dynamics}
    \dot{\mathbf{x}}_i = A\mathbf{x}_i + B\mathbf{u}_i, \ i \in \mathcal{A}
\end{equation}
where 
\begin{small}
\begin{equation}
    A = \left[
    \begin{array}{ccccc}
    0 & 1 & 0 & \ldots & 0 \\
    0 & 0 & 1 & \ldots & 0 \\
    \vdots & \ddots & \ddots & \ddots & \vdots \\
    0 & \ldots & \ldots & 0 & 1 \\
    0 & \ldots & \dots & 0 & 0 \\
    \end{array}
    \right] \otimes I_{m}, B = \left[
    \begin{array}{c}
    0 \\
    0 \\
    \vdots \\
    0 \\
    1
    \end{array}
    \right] \otimes I_{m}
\end{equation}
\end{small}
For the whole system, $\mathbf{x} = [\mathbf{x}_1^T, \mathbf{x}_2^T,...,\mathbf{x}_N^T]^T \in \mathbb{R} ^{Nnm}$. For UAVs $i,j \in \mathcal{A}$, $\mathbf{p}_{ij} = \mathbf{p}_i - \mathbf{p}_j$ is used to represent the relative position of UAV $i$ to UAV $j$. The neighbor set of UAV $i$ is denoted by $\mathcal{N}_i \subseteq \mathcal{A}$ and $\mathcal{N}_i \cap i = \emptyset$. There are many approaches to determine $\mathcal{N}_i$, \emph{e.g.}, the inter-UAV distance, $k$-nearest neighbors, and Voronoi partition \cite{fine_unifying_2013}. In this study, the $k$-nearest neighbors method is opted, which is also the case for some gregarious animals in nature \cite{Ballerini2008PNAS}. But with a slight change, the number of neighbors for each UAV is limited to reduce computation. Stipulate that the maximum number of neighboring UAVs is $k$. Meanwhile, if there is a leader in the group, its state is known globally, which means that the leader must become a neighbor of each UAV as shown in Fig \ref{fig:UAV_connection}.

Consider a group of UAVs moving freely in space. This group has a leader, denoted by "\emph{l}", and it can follow the desired path without being influenced by others. The other UAVs can observe and follow the leader's state. The problem we want to investigate is how to make the group exhibit flocking behavior while ensuring good performance, including efficiency, practical feasibility, \emph{etc.}

\section{Algorithm Design}\label{sec:Approach}
\subsection{State Prediction}\label{subsec:State Prediction}
Model predictive control (MPC) is capable of solving multi-constraint optimization and is regarded as a promising approach for flocking control by the advantage of a faster and smoother response \cite{lyu_multivehicle_2021}. This concept is also used in this paper and the inter-UAV interaction established by the potential function is a multi-constraint condition. Then the optimal control input can be obtained by solving the maximum probability as stated in Section \ref{subsec:Screening}.

\textbf{Control input discretization}: The control input space, denoted by $\bm{\mu}$, is firstly discretized into zero and non-zero. For the non-zero case, $\bm{\mu}$ is further divided by allowing discretized inputs uniformly distributed in a plane with the direction interval, $\theta_{min} = 2\pi/n_a$, where $n_a$ represents the number of nonzero control input's direction, $\theta_{min}$ is the angle between discretized adjacent non-zero control inputs. Hence, the control input discretization is denoted as $\bm{\mu} = \left\{\mathbf{0}, {u}_t\mathbf{e}_u(\theta_{min}),{u}_t\mathbf{e}_u \right.$ $\left.(2\theta_{min}), \ldots, {u}_t\mathbf{e}_u(n_a\theta_{min})\right\}$ in a plane, where $\mathbf{e}_u(x) = \left[\sin(x),\cos(x)\right]^T$, and ${u}_t$ is selected as an arithmetic progression with the interval $\bigtriangleup u$. This selection can be directly extended to the 3D case by simply changing the plane polar coordinates to the spatial spherical coordinates.
 
Based on equation \eqref{eq:dynamics}, the future state can be predicted for a selected control input $\mathbf{u}_t$ and $n=2$  after an Euler discretization with $t_p$.
\begin{equation}\label{eq:Eular prediction}
    \mathbf{x}_{t+t_p} = G\mathbf{x}_t + K\mathbf{u}_t
\end{equation}
where $\mathbf{x}_t$ and $t_p$ denote the current state and planning horizon, respectively, $G$ and $K$ are given as
\begin{small}
\begin{equation}
    G = \left[
    \begin{array}{cc}
    1 & t_p \\
    0 & 1
    \end{array}
    \right] \otimes I_{m}, K = \left[
    \begin{array}{c}
    t_p^2 / 2 \\ t_p
    \end{array}
    \right] \otimes I_{m}
\end{equation}
\end{small}
 
Hence, the predicted states after $t_p$ are $\mathbf{x}_{t+t_p}(\bm{\mu},\mathbf{x}_t,t_p)=\{\mathbf{x}_{t+t_p,1}(\mathbf{u}_1,\mathbf{x}_t,t_p),\mathbf{x}_{t+t_p,2}(\mathbf{u}_2,\mathbf{x}_t,t_p),\ldots\}$. To ensure the feasibility of the planned trajectory, it is necessary to explicitly impose constraints that limit the maximum control input and speed of each UAV, \emph{i.e.}, for $i \in \mathcal{A}$, $\Vert\mathbf{v}_i\Vert \leq v_{max}$ and $\Vert\mathbf{u}_i\Vert \leq u_{max}$.

\subsection{Potential function}\label{subsec:Potential function}
In this part, the method of selecting the form of potential function from how to form the collective behavior and improve the flight performance will be analyzed.

As heuristic flocking rules, cohesion and separation can ensure that multiple UAVs form a whole. And the equal and constant distance between the UAVs is hoped to keep when the group reaches the desired state. Thus the following function is chosen.
\begin{equation}\label{eq:attraction and repulsion}
    \varPsi_a(\mathbf{x}_i, \mathbf{x}_j) = -a \exp(-\frac{d_{ij}}{k_a}) + b \exp(-\frac{d_{ij}}{k_r})
\end{equation}
where $a$, $b$, $k_a$, and $k_r$ are all positive constants that determine the desired distance $d_t$ between UAVs, $ak_r < bk_a$ has to be satisfied so that there is a minimum, and $d_{ij} = \Vert \mathbf{p}_{ij} \Vert$. 

In fact, the gradient descent direction of the attraction-repulsion function is used to determine the control input. This is insufficient to meet practical employment needs, so the velocity alignment principle is discussed separately to highlight its importance. But not just the velocity direction, the energy between UAVs is also affected by its magnitude. A fast UAV travels a larger distance in a given amount of time (planning horizon $t_p$), which influences other flocking rules. Hence, the velocity alignment is chosen to be
\begin{equation}\label{eq:velocity alignment}
    \varPsi_{align}(\mathbf{x}_i, \mathbf{x}_j) = \exp(\frac{d_i\Delta \theta _{align}}{k_l})
\end{equation}
where $d_i = \Vert \mathbf{v}_i \Vert t_p$, and $\Delta \theta _{align} = \arccos (\mathbf{v}_i \cdot \mathbf{v}_j/ \left\Vert \mathbf{v} \right\Vert  \cdot \left\Vert \mathbf{v}_j \right\Vert)$, $j\in \mathcal{N}_i$.

As will be seen in section \ref{subsec:Screening}, the optimal control input is obtained iteratively based on $\bm{\mu}$. Increasing the dimension of $\bm{\mu}$ will raise the computational burden significantly, which is not feasible for online motion planning. This means that there is a large gap between each candidate control input, resulting in undesirable behaviors, including collision and insufficient execution. Therefore, the amplitude and direction of $\mathbf{u}$ are considered.
\begin{equation}\label{eq:acceleration cost}
    \varPsi_{acc}(\mathbf{x}) = \exp(\frac{\Vert \mathbf{u} \Vert}{k_c}) + \exp(\frac{\Delta \theta _{acc}}{k_d})
\end{equation}
where $\Delta \theta _{align} = \arccos(\mathbf{u} \cdot \mathbf{u}_{last} / \Vert \mathbf{u} \Vert \cdot \Vert \mathbf{u}_{last} \Vert)$, $\mathbf{u}_{last}$ is the previous control input.

No doubt adding equation \eqref{eq:acceleration cost} would smooth the velocity transition. However, it has no effect on position directly. Thus the velocity selection is improved, and the predicted velocity closest to the leader velocity $\mathbf{v}_l$ is more likely to be chosen. This is especially prominent when considering real dynamics.
\begin{equation}\label{eq:velocity cost}
    \varPsi_{vel}(\mathbf{x}) = \exp(\frac{\Vert\mathbf{v}-\mathbf{v}_l\Vert}{k_v})
\end{equation}

\subsection{Screening}\label{subsec:Screening}
The potential function, which offers an indicator for calculating the energy of the group states, was used to establish the UAV-UAV link. The multi-UAV system is now viewed as the MRF, where each UAV corresponds to a node in its probability graph $G$, and the system's energy distribution is stated as a joint probability distribution of random variables. Based on the joint probability distribution determined by the predicted states, the optimal control input can be selected.

Let $X = \left \{X_1, X_2,..., X_N \right \}$ represent a set of random variables, where $X_i$ corresponds to UAV $i$, with domain $\mathbf{x}_{t+t_p}(\bm{\mu}, \mathbf{x}_t, t_p)$. Then the joint probability density can be factored as $P(X) = 1/Z\exp{(-\sum _{\mathcal{Q} \in \mathcal{C}} \varPsi _{\mathcal{Q}}(X_{\mathcal{Q}}))}$, where $Z$ is a normalization factor, according to the set of cliques $\mathcal{C}$ in $G$ \cite{kindermann1980markov}. For each clique $\mathcal{Q}$, its energy would contain the following items.
\begin{subequations}
    \begin{equation}\label{eq:phi_acc}
    \phi_{acc}(X_i) = \exp(-\varPsi_{acc}(\mathbf{x}_i))
    \end{equation}
    \begin{equation}\label{eq:phi_vel}
    \phi_{vel}(X_i) = \exp(-\varPsi_{vel}(\mathbf{x}_i))
    \end{equation}
    \begin{equation}\label{eq:phi_a}
    \phi_a(X_i,X_j) = \exp(-\varPsi_a(\mathbf{x}_i, \mathbf{x}_j))
    \end{equation}
    \begin{equation}\label{eq:phi_align}
    \phi_{align}(X_i,X_j) = \exp(-\varPsi_{align}(\mathbf{x}_i, \mathbf{x}_j))
    \end{equation}
\end{subequations}
Thus the probability density distribution representing the energy distribution in the local region satisfies
\begin{equation}\label{eq:probdst}
    \begin{aligned}
	p(&X) =\frac{1}{Z} \exp\Bigg(-\varPsi_{acc}(\mathbf{x}_i) - \varPsi_{vel}(\mathbf{x}_i) \\
	& -\sum_{j\in\mathcal{N}_i} \varPsi_a(\mathbf{x}_i, \mathbf{x}_j) - \sum_{j\in\mathcal{N}_i} \varPsi_{align}(\mathbf{x}_i, \mathbf{x}_j)\Bigg)
    \end{aligned}
\end{equation}
	
However, it is difficult to determine $p(X)$ due to the unknown relationship between variables. We expect to seek an approximate distribution $q(X)$ to match $p(X)$ and minimize the KL divergence \cite{NIPS2011_beda24c1}. And by the fact that the predicted states of each UAV in the local area are influenced by its neighbors. This convergence process can be guaranteed through mutual iteration. Briefly, the update criterion can be obtained \cite{NIPS2011_beda24c1}.
\begin{equation}\label{eq:criterion}
    \begin{aligned}
    Q_i(&\mathbf{x}_{t+t_p, i}) = \\
    &\frac{1}{Z_i} \exp \Bigg(\sum_{j\in\mathcal{N}_i} Q_j(\mathbf{x}_{t+t_p, j}) \varPsi_a(\mathbf{x}_{t+t_p, i}, \mathbf{x}_{t+t_p,j})  \\
    &- \sum_{j\in\mathcal{N}_i} Q_j(\mathbf{x}_{t+t_p, j}) \varPsi_{align}(\mathbf{x}_{t+t_p, i}, \mathbf{x}_{t+t_p, j}) \\
        &- \varPsi_{acc}(\mathbf{x}_{t+t_p, i}) - \varPsi_{vel}(\mathbf{x}_{t+t_p, i})\Bigg)
    \end{aligned}
\end{equation}
where, the initial condition can be $Q_i(\mathbf{x}_{t+t_p, i}) = 1 / number$ $(\mathbf{x}_{t+t_p, i}(\bm{\mu}, \mathbf{x}_{t+t_p, i}, t_p))$.

\subsection{Smoothing}
In section \ref{subsec:State Prediction}, $n=2$ is inconsistent with actual physical systems. The immediate consequence is that the UAV cannot accurately follow the desired command. Hence, low-pass filtering is introduced.
\begin{equation}\label{eq:filter}
    \mathbf{u}^* = (1 - \alpha)\mathbf{u}_{last} + \alpha\mathbf{u}
\end{equation}
where $0 < \alpha < 1$. Reducing $\alpha$ slightly will have the same effect as equation \eqref{eq:acceleration cost}.
\begin{figure*}
    \centering
    \includegraphics[width=0.96\linewidth]{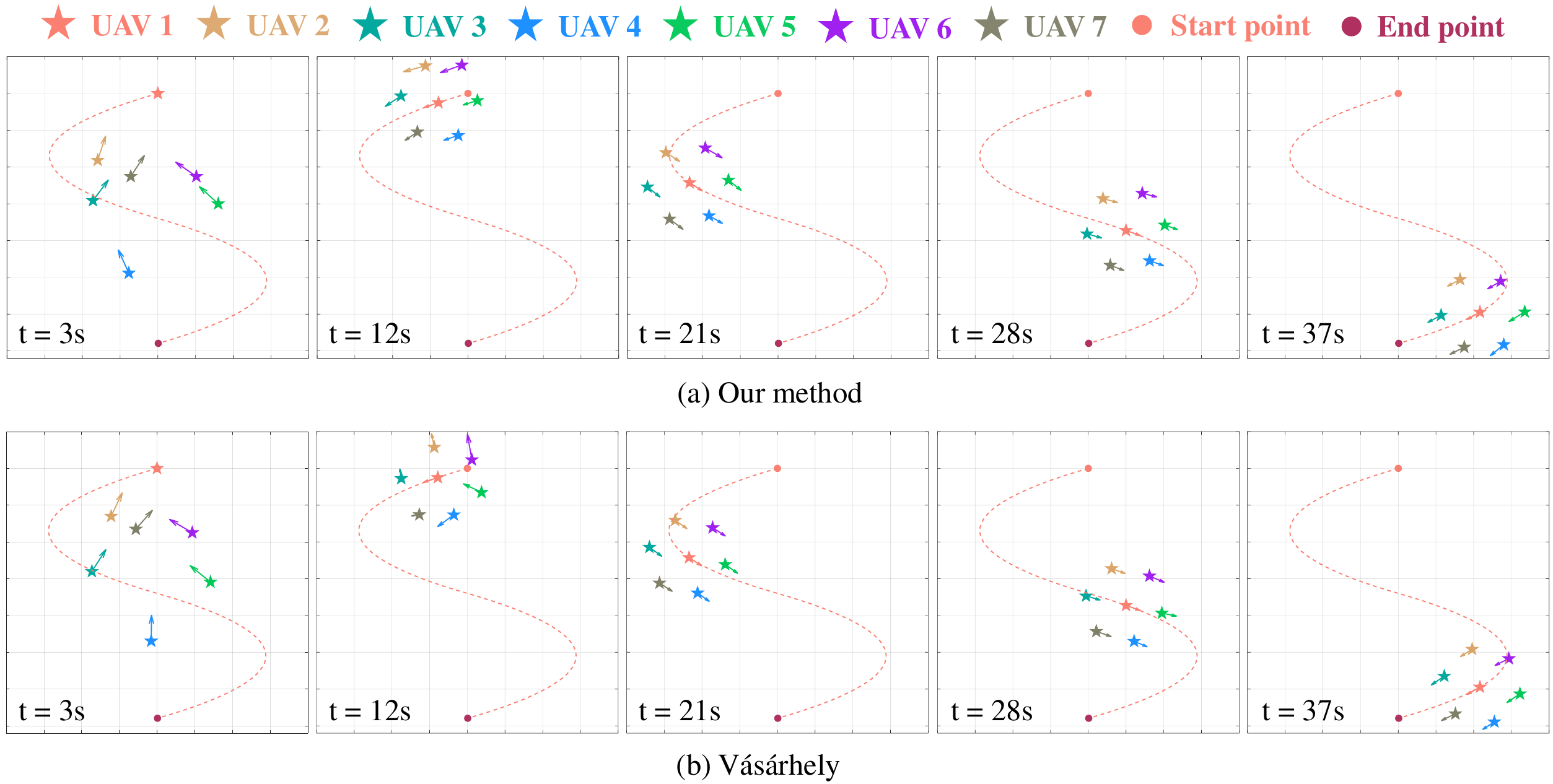}
    \caption{Simulation trajectory. Both methods can ensure that the UAV swarm is distributed in a grid shape and the velocity is basically consistent. And the proposed method can reach the desired state faster (the difference is most obvious at $t=12$ $\mathrm{s}$).}
    \label{fig:simulation_trajectory}
\end{figure*}

\begin{table}[!htb]
    \centering
    \caption{Parameters in simulation and experiment}
    \label{tab:parameters}
    \setlength{\tabcolsep}{2em}{
    \renewcommand\arraystretch{1.1}
    \begin{tabular}{c|c|c}
      \hhline
      Parameters & simulation & experiment \\ \hhline
      $k$ & 3 & 2 \\ \hline
      $a$ & 8 & 8 \\ \hline
      $b$ & 10 & 10 \\ \hline
      $k_a$ & 1.5 & 1.5 \\ \hline
      $k_r$ & 0.2 & 0.27 \\ \hline
      $n_a$ & 6 & 6 \\ \hline
      $\bigtriangleup u$ & 0.14 & 0.14 \\ \hline
      $\triangle t(s)$ & 0.05 & 0.02 \\ \hline
      $t_p(s)$ & 0.15 & 0.1 \\ \hline
      $k_l$ & 4 & 4 \\ \hline
      $k_c$ & 7 & 7 \\ \hline
      $k_d$ & 15 & 2 \\ \hline
      $k_v$ & 2 & 1 \\ \hline
      $\Vert\mathbf{v}_l\Vert(\mathrm{m/s})$ & 0.2 & 0.12 \\ \hline
      $\alpha$ & 0.8 & 0.9 \\ \hline
      $r_{coll}(\mathrm{m})$ & 0.12 & 0.1 \\
      \hhline
    \end{tabular}}
\end{table}
\section{Simulation and Experiment}\label{sec:Experiment}
To evaluate the flight performance, some common metrics are introduced.
	
The \textbf{order metric} describes the velocity correlation, which corresponds to the velocity alignment term.
\begin{equation}\label{eq:order metric}
    order = \frac{1}{N} \sum_{i \in \mathcal{A}} \frac{1}{N_i - 1}\sum_{j \in \mathcal{N}_i} \frac{\mathbf{v}_i \cdot \mathbf{v}_j}{\Vert\mathbf{v}_i\Vert \cdot \Vert\mathbf{v}_j\Vert}
\end{equation}
where $N_i=number(\mathcal{N}_i) + 1$. The desired flocking state is that the order metric is close to 1.
	
The \textbf{distance metric} calculates the distance between each UAV and its nearest neighbor. The minimum, maximum, and average separations among UAVs are used as the distance metrics, which are defined as
 \begin{equation}\label{eq:distance}
    \begin{aligned}
        d_i^{min} &= \min\left\{d_{ij} \vert d_{ij}=\|\mathbf{p}_{ij}\|, \; \forall j\in\mathcal{A}, j\neq i\right\} \\
    d^{min} &= \min\left\{d_{i}^{min} \vert \; \forall i\in\mathcal{A}\right\} \\
     d^{max} &= \max\left\{d_{i}^{min} \vert \; \forall i\in\mathcal{A}\right\} \\
    d^{avg} &= \mathrm{mean}\left\{d_{i}^{min} \vert \; \forall i\in\mathcal{A}\right\} \\
    \end{aligned}
\end{equation}

The \textbf{control efficiency metric} evaluates the input efficiency of each UAV, which is defined as  
\begin{equation}\label{eq:mean control input}
    u_i^{avg} = \frac{1}{T} \sum_{t=0}^{T} \Vert\mathbf{u}_i(t)\Vert
\end{equation}
where $T$ is the total running time.
 
The \textbf{trajectory length metric}: a short trajectory length is preferable.
\begin{equation}\label{eq:trajectory length}
    L_i = \sum_{\substack{t=0 }} ^{T}\Vert\mathbf{p}_i(t+\triangle t) - \mathbf{p}_i(t)\Vert
\end{equation}
where $\triangle t$ is the step size in simulation and experiments.
\begin{figure}[!htbp]
    \centering
    \includegraphics[width=1\linewidth]{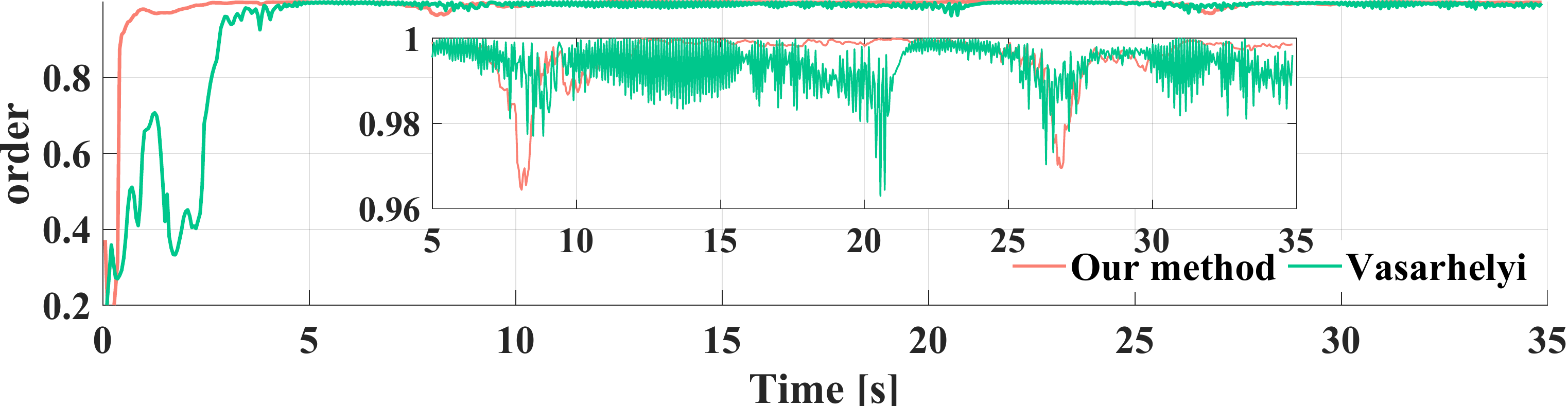}
    \caption{$order$. The order implemented with the proposed method can be guaranteed to be close to 1 and recover quickly after a disturbance. However, the reaction of Vásárhelyi's method is slower, and there are even small fluctuations.}
    \label{fig:simulation_order}
\end{figure}
\begin{figure}[!htbp]
    \centering
    \includegraphics[width=1\linewidth]{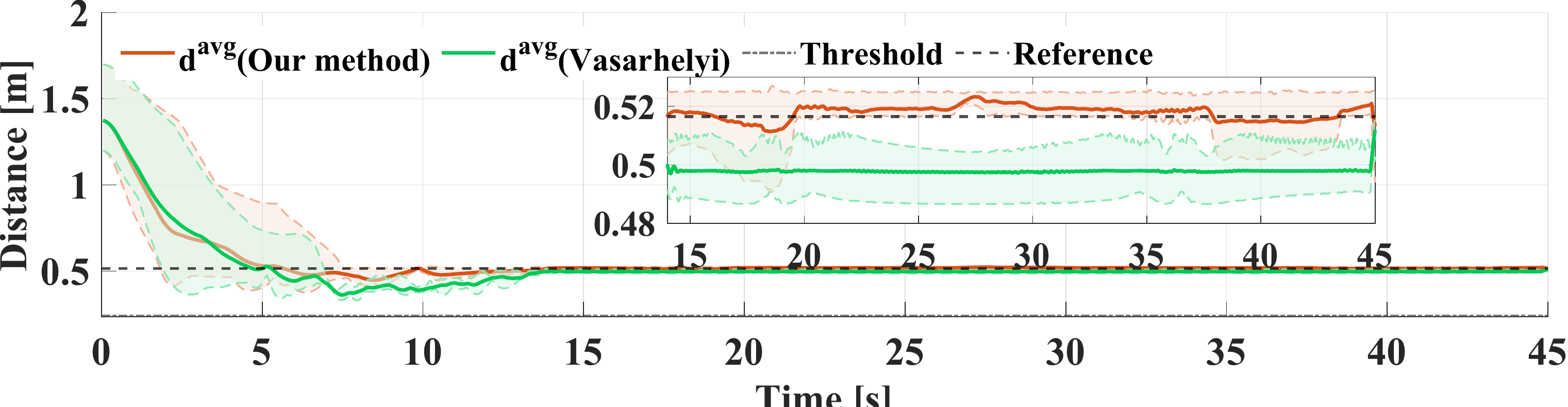}
    \caption{$d^{min},d^{max},d^{avg}$. For a given desired distance $d_t=0.51 63$ $\mathrm{m}$, both methods can avoid collisions, $d_{min} \geq 0.24$ $\mathrm{m}$. But the proposed approach has better distance retention with a smaller error.}
    \label{fig:simulation_distance}
\end{figure}
\begin{figure}[!htbp]
    \centering
    \includegraphics[width=1\linewidth]{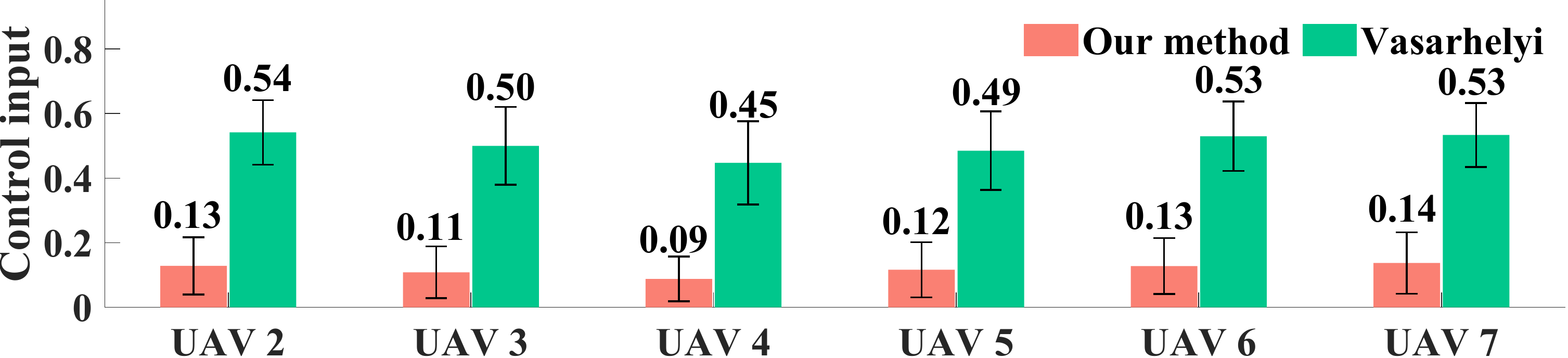}
    \caption{$u^{avg}$. The proposed method explicitly imposes constraints on the acceleration, and the control efficiency is about 4.3 times that of Vásárhelyi's method.}
    \label{fig:simulation_control_input}
\end{figure}
\begin{figure}[!htbp]
    \centering
    \includegraphics[width=1\linewidth]{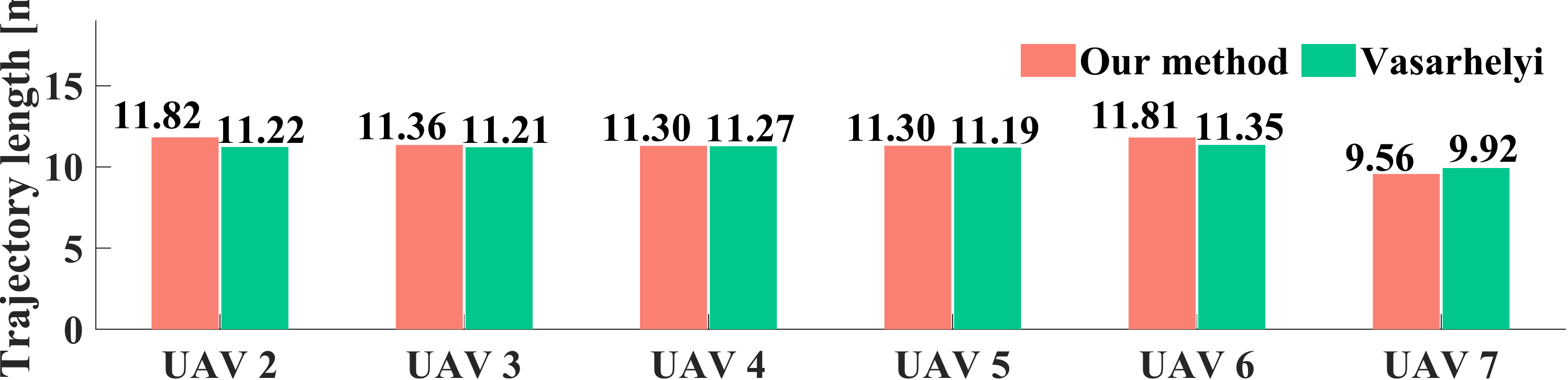}
    \caption{$L$. The V\'{a}s\'{a}rhelyi's method has a slightly smaller trajectory length than our method, which is one of its advantages.}
    \label{fig:simulation_trajectory_length}
\end{figure}

\subsection{Simulation}
The proposed algorithm is evaluated in a typical scenario. There are 7 UAVs and UAV 1 is the leader. Fig. \ref{fig:real_trajectory},\ref{fig:simulation_trajectory} depicts the procedure, where the UAVs are far apart in the beginning, and no flocking occurs. Following that, each UAV advances toward the leader to form a flock, and the leader moves along the desired S-shaped trajectory while the others follow to guarantee flocking behavior. Set $k=3$, $u_{max} = 0.7$ $\mathrm{m/s^2}$, and $v_{max} = 0.35$ $\mathrm{m/s}$. Other parameters are illustrated in Table \ref{tab:parameters}. The proposed algorithm is implemented in MATLAB. To demonstrate the efficiency of our algorithm, the proposed method is compared with V\'{a}s\'{a}rhely's \cite{Vasarhelyi2018SR}. Note that the neighbor choices in Vásárhelyi's algorithm follow the same routine as ours.

As illustrated in Fig. \ref{fig:simulation_trajectory}-\ref{fig:simulation_trajectory_length}, the flight performance implemented by the proposed algorithm outperforms that of the Vásárhelyi's in many ways, including group velocity correlation, distance retention ability, control efficiency, reactivity. The proposed method chooses an optimal control input from the predicted states to minimize the sum of all energies. Because the costs are considered separately, it is simple to ensure that group performance is close to expectations. Instead, Vásárhelyi's method is built around a velocity alignment mechanism that considers acceleration constraints. Some parameters, such as repulsion gain $p^{rep}$ and gain of braking curve $p^{frict}$, are sensitive and not only have a direct impact on distance but also may cause group shock. Furthermore, there are no other constraints to improve group performance, ensuring superiority in all aspects is difficult.
\begin{figure}
    \centering
    \includegraphics[width=1\linewidth]{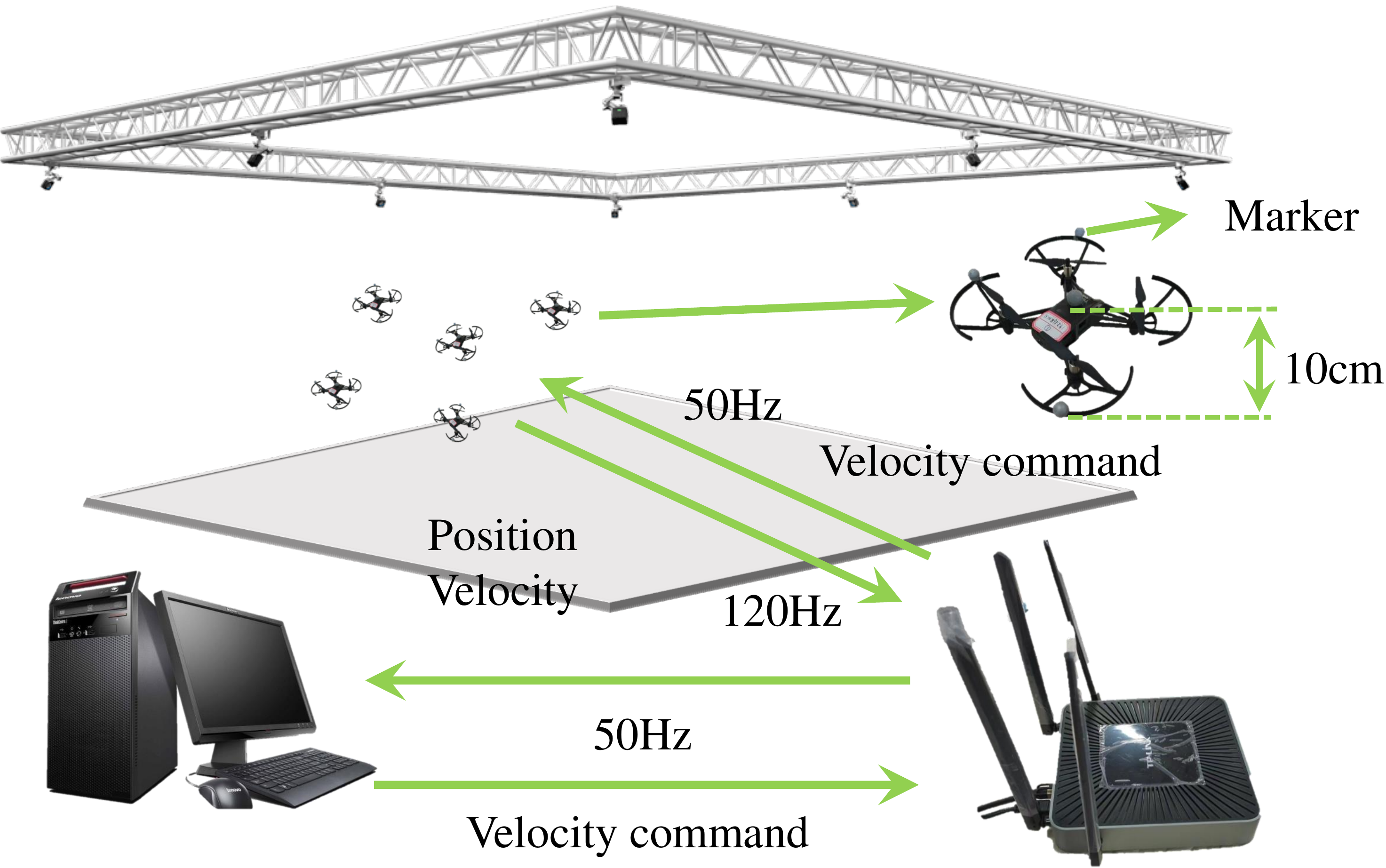}
    \caption{Hardware communication. Optitrack could transmit Tello's position and speed information at 120 Hz. The computer will receive position and speed information by setting $\triangle t=0.02$ $\mathrm{s}$. Tello accepts velocity commands to realize flocking behavior.}
     \label{fig:real_communication}
\end{figure}

\begin{figure}[!htbp]
    \centering
    \includegraphics[width=1\linewidth]{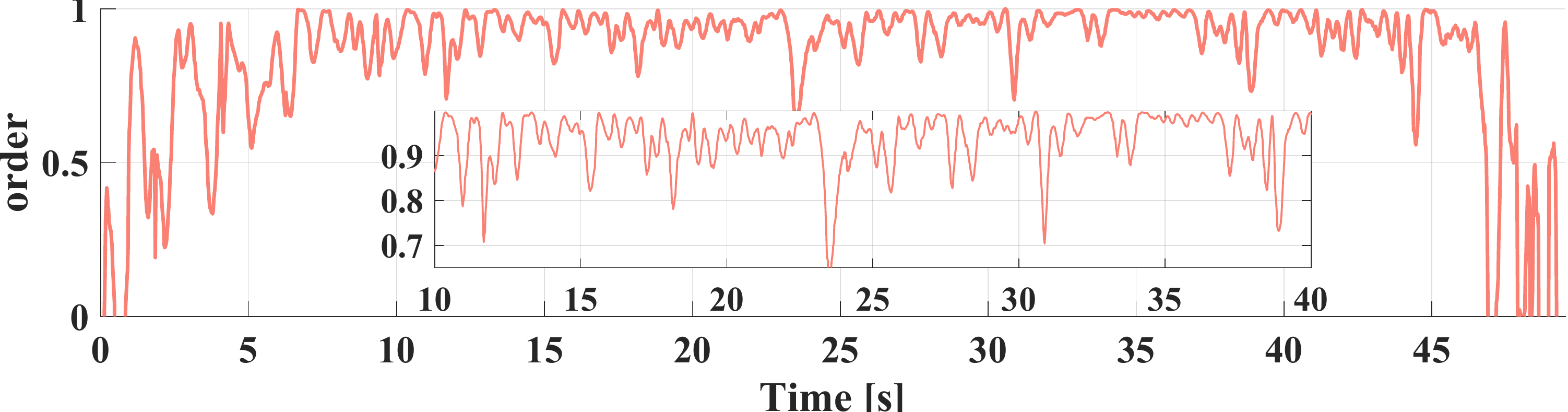}
    \caption{In the real experiment, except for some vibration in the middle process and the change of motion state at the beginning and end, the consistency of the UAVs can be basically maintained above 0.8.}
    \label{fig:real_order}
\end{figure}
\begin{figure}[!htbp]
    \centering
    \includegraphics[width=1\linewidth]{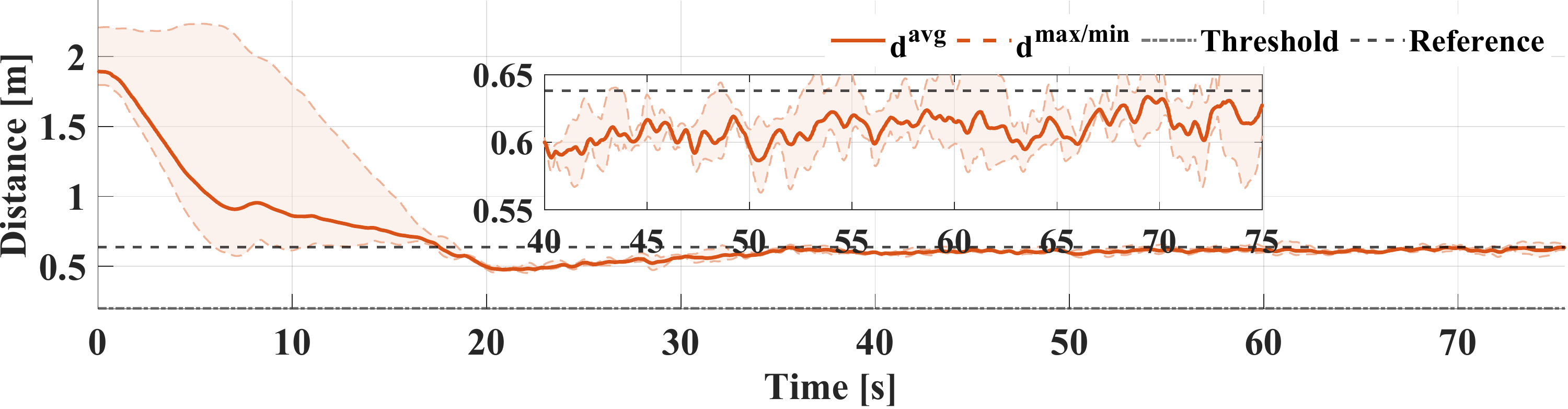}
    \caption{Despite inertia, delay, and some uncertainties, the distance between the UAVs is close to the desired distance $d_t=0.6381$ $\mathrm{m}$, while maintaining safety $d_{min} \geq 0.2$ $\mathrm{m}$.}
    \label{fig:real_distance}
\end{figure}
\begin{figure}[!htbp]
    \centering
    \includegraphics[width=1\linewidth]{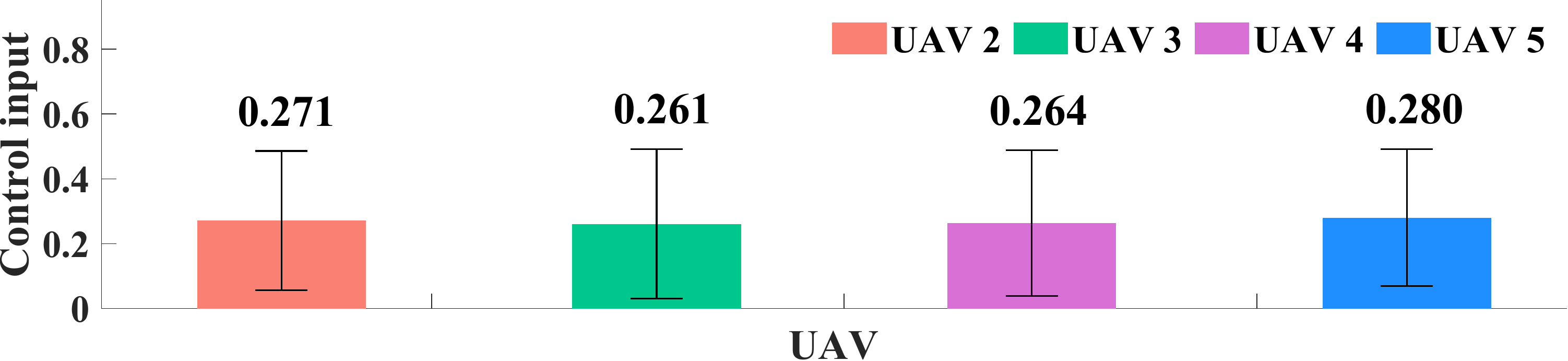}
    \caption{The average control input of the follower UAVs, $0.2690 \pm 0.2203$ $\mathrm{m/s^2}$(All follower UAVs)}
    \label{fig:real_control_input}
\end{figure}
\subsection{Experiment}\label{subsec:Real Experiment}
Real-world experiments are also performed to validate the proposed algorithm. The experimental scene is similar to the one in the simulation (Fig. \ref{fig:real_trajectory}). And 5 DJI Tello drones are used, whose states are obtained by the OptiTrack system as shown in Fig. \ref{fig:real_communication}. The Tello drones can track velocity commands using the implemented inner-loop controller. No interface is provided for the acceptance of acceleration commands, so the commands are modified to the velocity ones.
\begin{equation}\label{eq:velocity command}
    \mathbf{v}^*=\mathbf{v}+ \mathbf{u}^* \triangle t
\end{equation}

\begin{figure}[!htbp]
    \centering
    \includegraphics[width=0.99\linewidth]{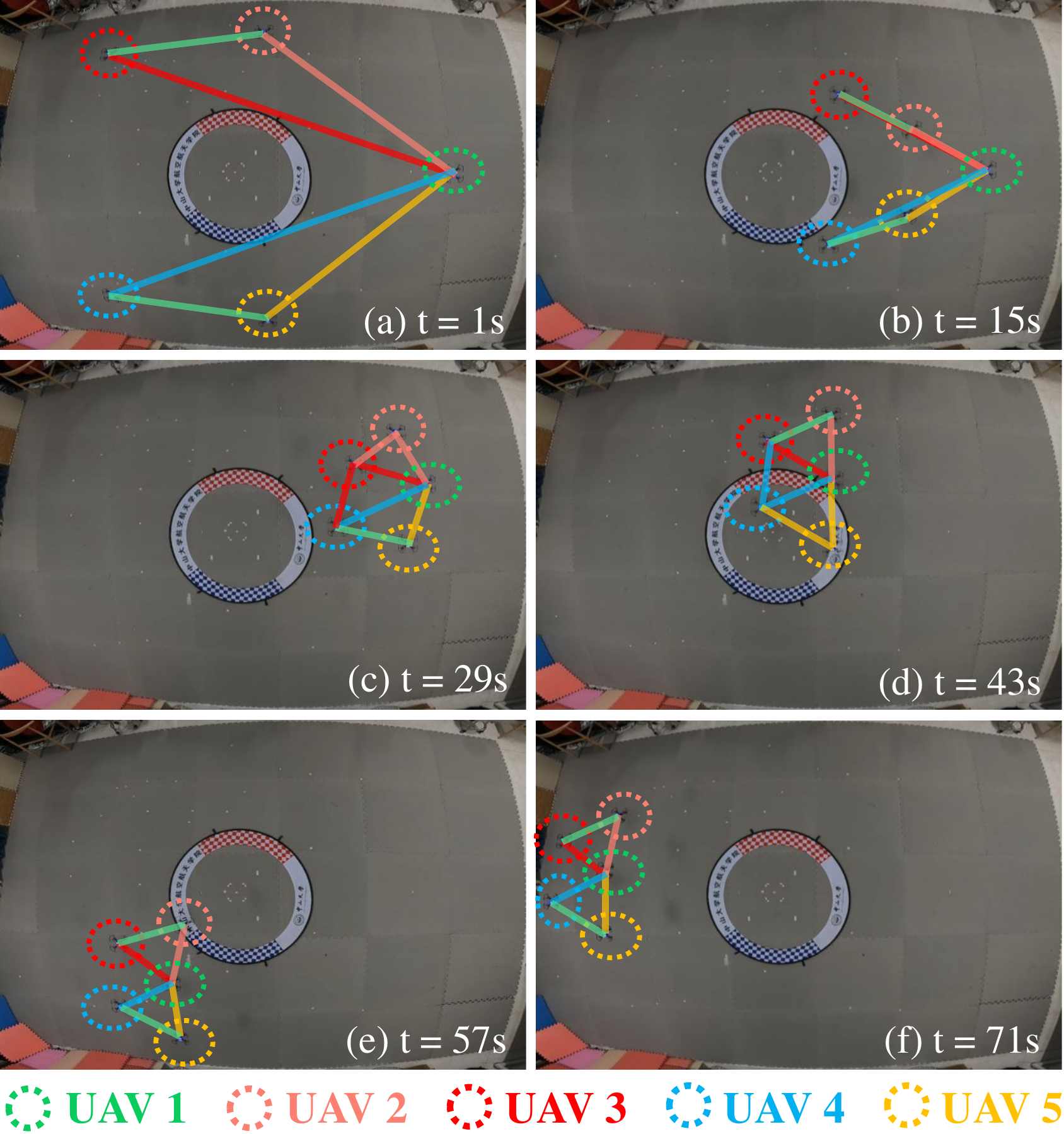}
    \caption{Experiment snapshots. UAVs can receive information from a neighbor if they are connected by a line with the same color. The light green line indicates that the two UAVs can communicate with each other.}
    \label{fig:real_snapshots}
\end{figure}

Due to external factors, including the instability of the Optitrack's data, the inertia of UAVs, data transmission delay, \emph{etc.}, the drones have a time constant of $0.3 \sim 0.4$ $\mathrm{s}$ and a response delay of $0.2 \sim 0.4$ $\mathrm{s}$, which poses a significant challenge for algorithm verification. These issues are mitigated by appropriately increasing the number of forward-predicting steps. Some key parameters are listed in Table \ref{tab:parameters}. The results show that the direction of group velocity is roughly the same, and the distance whose error from the expected value is only $3$ $\mathrm{cm}$ between drones is relatively stable. In addition, the control input has a raise to improve the response. However, relative to $0.7$ $\mathrm{m/s^2}$, control input and velocity costs still have a significant effect. Fig. \ref{fig:real_snapshots} shows the communication. Because each follower counts the leader as its neighbor, there is a straight line of color similar to itself connected to the leader. The communication between followers is related to the distance and $k$. It can be seen that after surrounding the leader, 4 followers keep the same distance from UAV 1 and move along the desired trajectory, which verifies the feasibility of the algorithm.

\section{Conclusion}\label{sec:Conclusion}
In this paper, the problem of forming a flock while following the desired trajectory was solved. Unlike many flocking control methods with passive characteristics, the idea of MPC was adapted to generate a set of candidate feasible trajectories according to the UAV dynamic model. The corresponding costs for flight performance, from the basic characteristics of the flocks to group performance, were considered. Especially, the control input and velocity costs had a significant effect on improving velocity correlation and control efficiency. Screening out the optimal control input was converted to solve the joint probability distribution of the MRF by an updated criterion. Compared with the existing Vásárhelyi's method, the effectiveness and feasibility of the proposed algorithm were demonstrated via simulation and experimental results.

\bibliographystyle{gbt7714-numerical}
\bibliography{myref.bib}

\end{document}